\documentclass[a4paper,12pt]{article}
\usepackage{graphics}
\usepackage{graphicx}
\usepackage{amsmath}

\title{Stationary shear flows of dense granular materials :
a tentative continuum modelling} 

\author{C. Josserand, P.-Y. Lagr\'ee and D. Lhuillier\\
 Laboratoire de Mod\'elisation en M\'ecanique,\\ Universit\'e P. et M. 
Curie (Paris 6) et CNRS \\
Case 162, 4 place Jussieu, 75252 Paris cedex 05, France
} 

\begin{document}

\maketitle

\begin{abstract}
We propose a simple continuum model to interpret the shearing motion of dense,
dry and cohesion-less granular media. Compressibility, dilatancy and 
Coulomb-like friction are the three basic ingredients. The granular stress is 
split into a rate-dependent part representing the rebound-less impacts 
between grains and a rate-independent part associated with long-lived contacts.
Because we consider stationary flows only, the grain compaction and the grain 
velocity are the two main variables. The 
predicted velocity and compaction profiles are in apparent agreement with the 
experimental or numerical results concerning free-surface shear flows as well as confined shear flows
\end{abstract}

\section{Introduction}

The mechanical behaviour of a flowing granular material depends strongly on the grain compaction. While dense granular media usually exhibit relatively 
slow motions with predominance of friction, less dense ones are usually 
found in vigorous motions with predominance of two-particles collisions. 
The collision-dominated regime is well described by kinetic theory, with the 
concepts of granular temperature and inelastic collisions. On the contrary, 
the current description of dense granular flows is not so fully satisfactory. 
It must be understood that we are not questioning the description by soil 
mechanics of quasi-static and highly stressed granular materials, but the 
description of flows with relatively low stress levels encountered, for 
example, in avalanches down an inclined plane.
Several recent works (see e.g. \cite{azanza,raj03,anceve}) concluded 
rather pessimistically about the possibility of describing dense granular 
flows within the realm of continuum mechanics. In fact, the experimental 
observation that most dense flows display a typical thickness of a few grain 
diameters must not be a factor of pessimism. We know from several examples in 
suspension mechanics that the continuum approach can cope with high velocity 
gradients in one direction, provided one has some statistical homogeneity in 
the other two directions. This situation is exactly the one met in sheared 
granular media, provided we discard transient effects and focus on the final 
stationary state. Once the continuum description is accepted, the number of 
relevant field variables must be decided. There is no doubt that the grain 
velocity is relevant and it is not less clear that the grain compaction is 
also a pertinent variable. In fact the widely used assumption of an 
incompressible medium is not tenable. It contradicts the dilatancy concept 
and, as will be seen below, the transport coefficients of a dense granular 
medium display enormous variations with only tiny modifications of the 
compaction. Our aim is thus to propose a model for dense and 
stationary shear flows in which the grain compaction and the grain velocity 
are the two fundamental variables. One could also suggest the fluctuational 
kinetic energy of the grains (the granular "temperature") as a third variable. 
However, since we limit our analysis to stationary shears, the granular 
temperature is no longer an independent variable. The role of the embedding 
fluid will be neglected everywhere, and for these "dry" granular media, the 
main issue is to propose a constitutive relation for the granular stress.
   To compare with previous works on dense flows, we can say we adopt a 
phenomenological description somewhat similar to that proposed two decades 
ago by Savage \cite{savage82} and by Johnson and Jackson \cite{johnson}. Like these 
authors, we introduce a stress tensor split into a frictional and a collisional 
contribution. However, the collisional contribution is concerned with rebound-less impacts characteristic of high grain concentration, and is free of any restitution coefficient. Our constitutive relation for the 
particulate stress has a form somewhat similar to that proposed by Ancey and Evesque \cite{ancey}, the main differences concerning the role of the grain compaction and a more detailed expression of the granular pressure. Our 
model also shares some common features with the model proposed by Bocquet 
et al. \cite{bocquet} and by Louge \cite {louge}, 
but instead of extending the kinetic theory approach to large compaction, we prefer here to develop a model specifically devoted to dense media.
Moreover, the model we propose is quite simple in so 
far as it denies any special role to the compaction gradient \cite{goco79} and avoids the non-locality concept \cite{mills99}. 

Discarding two-particles 
collisions and any restitution coefficient means that our model is restricted to 
volume fractions in the range between $\phi_m$ and $\phi_M$. The maximum 
grain compaction $\phi_M$ corresponds to the highest possible random packing 
(with $\phi_M \simeq 0.80$ for two-dimensional flows and $\phi_M \simeq 0.65$ 
for three-dimensional ones) while $\phi_m$ is the smallest compaction 
compatible with the existence of a continuous network of contacts between 
grains. As suggested by Azanza \cite{azanza}, one can define $\phi_m $
as the minimum compaction for which the two-particle distribution function 
exhibits some swelling at a distance of {\it two} diameters. With this definition, 
$\phi_m \simeq 0.70 $ for two-dimensional flows while $\phi_m \simeq
0.50 $
for three-dimensional ones.

A phenomenological order parameter description of granular media was recently 
proposed \cite{arts01}. We acknowledge this approach looks efficient in 
describing a large number of phenomena observed in dense flows, but it suffers 
from two serious drawbacks. The first one is the huge number of possible 
candidates for the order parameter. For example, in the model to be presented below, the 
reduced compaction $(\phi-\phi_m)/(\phi_M-\phi_m)$ could play this role. A 
second difficulty concerns the relevance of the Ginzburg-Landau relaxation 
equation for the order parameter. Here we consider the standard conservation 
of mass and momentum only, without any extra equation.

The description of stationary free-surface shear flows is given in section \ref{sex:free} 
while that of confined shear flows is presented in section \ref{sex:con}. 
Section \ref{sex:exp} compares the model predictions with experimental and (or)
numerical data. The final 
section insists on the limitations and necessary improvements of the proposed 
model, which must be considered as a minimal one.

\section{Free surface shear-flows}\label{sex:free}

\begin{figure}
\centerline {
\includegraphics[width=10cm]{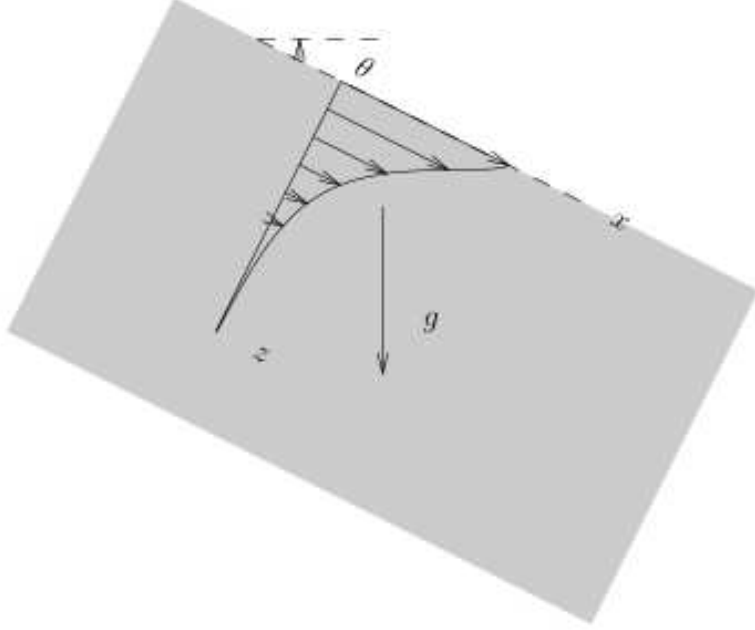}}
\caption{  
Gravity-induced  shear flow with free surface (over a heap or an inclined plate  with an angle 
$\theta$ relative to the horizontal plane).} 
\label{fig:fig1}
\end{figure}

As a prototype of shear flow with free surface, we consider the 
gravity-induced chute (over a heap or an inclined plate, see Figure \ref{fig:fig1}) with an angle 
$\theta$ relative to the horizontal plane. 
The mean grain velocity is parallel to the $x$-axis, 
${\bf V} = V {\bf e}_x$, while $V$ and the solid fraction $\phi$ depend only on 
$z$, the {\it distance to the free surface}. The granular stress tensor is 
noted $\tau$ and the equations of motion are:

\begin{equation}
0  =  -\frac{\partial \tau_{xz}}{\partial z}  +  \phi \rho g {\rm sin}(\theta)\; , \;\;\;\;
0  =  -\frac{\partial \tau_{zz}}{ \partial z}  + \phi \rho g {\rm cos}(\theta) 
\label{stress}
\end{equation}
\noindent
where $\rho$ is the constant mass per unit volume of the grain material and $g$ is the acceleration of gravity.
 
For dense granular media, the granular stress is a consequence of long-lived
contacts and bounce-less impacts between grains. Long-lived contacts result from compressive forces acting towards the boundaries of the granular medium. In the geometry considered, they take part in $\tau_{zz}$ since $z$ is the direction of main compression. Whether gravity is responsible for compressive forces or not, we choose to scale the compressive stress with $\rho gD$ where $D$ is the grain size. The compressive stresses are related to the grain volume fraction as $\rho gD F(\phi)$, where $dF/d\phi$ is the non-dimensional compressibility of the granular medium. In free-surface shear flows, gravity is the only source of compaction and the magnitude of the compressive stress will also depend on $\theta$. It is clear that the compressive role of gravity is maximum when the compression axis $z$ is vertical while this role vanishes when gravity is orthogonal to it. Consequently, the general form of the gravity-induced compressive stress is $\rho gD F(\phi)f(\theta)$ with $f(0)=1$ and $f(\pi/2)$= 0. The exact expression of $f(\theta)$ is not important because, as will soon be seen, the stationary flows exist in a very limited range of $\theta$ only. One of the simplest function of $\theta$ which meets the above requirement is $\cos(\theta)$, and we assume henceforth that the contribution of long-lived contacts to $\tau_{zz}$ can be written in the form $\rho gD F(\phi)cos(\theta)$. To this gravity-induced contact stress must be added a 
rate-dependent impact stress. On purely dimensional grounds, this second 
contribution cannot be but Bagnold-like and the full normal stress finally 
appears in the form:
\begin{equation}
\tau_{zz}  =   \rho D^2 \mu_N(\phi) \left(\frac{dV}{dz} \right)^2  +\rho g D F(\phi)  {\rm cos} (\theta),
\label{tzz}
\end{equation}
\noindent
where $\mu_N(\phi)$ represents the compaction-dependent intensity of the normal 
stress induced by the shear rate. Concerning  the shear stress of the flowing 
granular medium, we assume it is 
made of a Coulomb-like contribution with a friction coefficient $\mu(\phi)$
completed by a Bagnold-like contribution involving a coefficient $\mu_T(\phi)$
representing the compaction-dependent intensity of the shear stress induced 
by the shear rate
\begin{equation}
\tau_{xz} = \rho D^2 \mu_T(\phi) \left(\frac{dV}{dz} \right)^2+\mu(\phi) \tau_{zz},
\label{txz}
\end{equation}
The model expressions (\ref{tzz}) and (\ref{txz}) contain four functions of 
the grain compaction. Before giving them some explicit (and tentative) 
expressions, let us comment on their expected general behaviour. These four 
functions are characteristic of the dense regime and have a meaning in the 
range $\phi_m \le \phi \le \phi_M$ only. We expect $F$, $\mu_T$ and $\mu_N$ to 
become infinite when $\phi= \phi_M$, because no 
motion nor extra compaction is expected above the maximum random packing. We 
also expect $F$ and $\mu_N$ to vanish for $\phi= \phi_m$, because the normal 
stresses must vanish for the most tenuous contact network. Concerning the 
friction coefficient $\mu$, it is the 
only coefficient which remains finite when $\phi=\phi_M$ and it presumably increases
\cite{chevoir} for 
smaller compactions.  
In short, the three scalars $F$, $\mu_N$ and $\mu_T$ are strongly increasing 
functions of the compaction, while $\mu$ has a much smoother behaviour.

Since we neglect the role of the embedding fluid, the granular stress must 
vanish at the free surface and consequently $\tau_{xz}=\tan(\theta) \tau_{zz}$
everywhere. In this case, when solving the equations of 
motion (\ref{stress}) with the model expressions (\ref{tzz}) and (\ref{txz}), 
one arrives at a compaction profile and a velocity profile which are solution 
of:
\begin{equation}
D \frac{d\phi}{dz}  =  \frac{\phi}{
\frac{\partial}{\partial \phi}    
[ \frac{F}{1 - (\mu_N/\mu_T)(\tan(\theta) - \mu)} ]
}
\label{dens}
\end{equation}
and
\begin{equation}
\left( \frac{D}{g} \right)^{1/2} \frac{dV}{dz}  =
  -\left(  \frac {F( {\rm sin}(\theta) - \mu  {\rm cos}(\theta)) } {\mu_T (1 - (\mu_N/\mu_T)(\tan(\theta) - \mu))}   \right)^{1/2}.
\label{velo}
\end{equation}
At the free-surface the solid fraction is $\phi_m$ (remember we limit the
description to the dense regime). According to (\ref{dens}) the solid fraction 
increases towards its maximum value $\phi_M$ over a depth which 
scales with the grain diameter but depends on $\theta $ if $\mu_N/\mu_T$ is 
different from zero. Hence $\mu_N/\mu_T$ represents the relative magnitude 
of Reynold's dilatancy. 
Concerning the velocity profile, its characteristic value scales like 
$(gD)^{1/2}$ and according to (\ref{velo}) its solution exists for any 
angle $\theta$ verifying the inequality  
$ \mu(\phi) \le \tan(\theta) \le \mu(\phi) + \mu_T(\phi)/\mu_N(\phi)$.
For certain values of $\theta$ this 
inequality is possibly satisfied in a part only of the full range $\phi_m \le
\phi \le  \phi_M$.

It is obviously not evident to deduce four functions of the compaction from 
the rather scarce experimental or numerical results on stationary shear flows. 
{\it We assume henceforth that} $\mu$ and 
$\mu_T/\mu_N$ {\it are independent of the grain compaction}. 
Then, a stationary solution is possible 
in a well-defined angle range $ \theta_{min} \le \theta \le \theta_{max} $, with $\tan(\theta_{min})= \mu$ and $ \tan(\theta_{max}) = \mu + \mu_T/\mu_N$. 
To obtain more quantitative results we consider separately the chute over a 
heap from that over an inclined plane.

\subsection{Heap flows}

\begin{figure}
\centerline {\includegraphics[width=10cm]{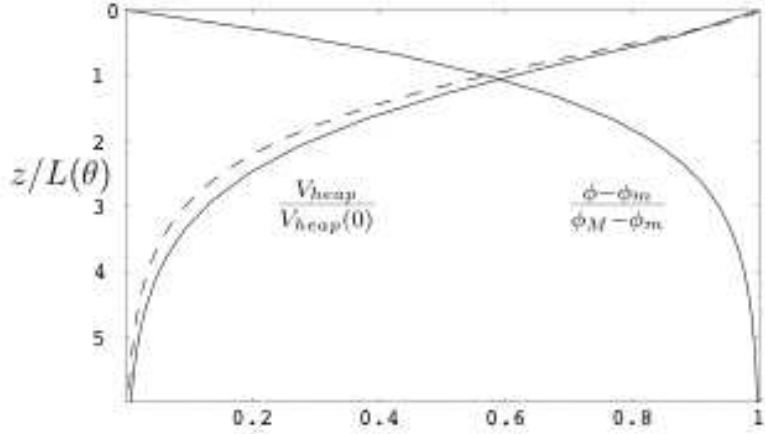}}
\caption{ Reduced velocity profile $V_{heap}/V_{heap}(0) $ versus the adimensional distance
$z/L(\theta)$ to the free surface. The dashed curve represents the approximate expression (\ref{vsimpl}). 
The reduced compaction profile $(\phi -\phi_m)/(\phi_M -\phi_m)$ is plotted as well.} 
\label{fig:fig2}
\end{figure}

\begin{figure}
\centerline {  
\includegraphics[width=10cm]{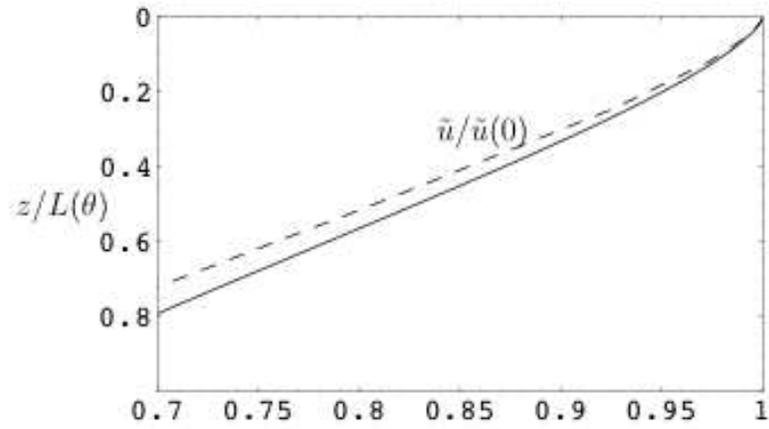}}
\caption{Zoom of figure \ref{fig:fig2} showing the Bagnold like region for $z/L(\theta)<0.2$. The dashed curve represents  approximation (\ref{vsimpl}). }
\label{fig:fig3}
\end{figure}

\begin{figure}
\centerline {
\includegraphics[width=10cm]{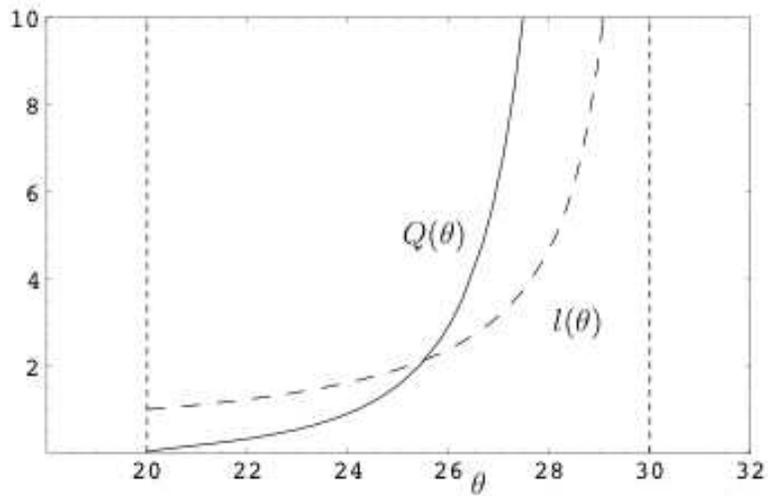}}
\caption{ $\theta$-dependance of  flux  $ Q_{heap} $ and 
adimensional  thickness $l(\theta)=\phi_M L(\theta)/D$. } 
\label{fig:fig4}
\end{figure}

In the heap case, provided $\mu$ and $\mu_N/\mu_T$ are independant of the
solid fraction, one can deduce from (\ref{dens}) and (\ref{velo}) the total
 granular flux flowing down the heap $Q_{heap}$:
\begin{equation}\label{qheap}
\frac{Q_{heap}}{D\sqrt{gD}}=\frac{( {\rm sin}(\theta)-\mu  {\rm cos}(\theta))^{1/2}}
{\left(1-\frac{\mu_N}{\mu_T}(\tan(\theta)-\mu)\right)^{5/2}} 
\int_{\phi_m}^{\phi_M} \left(\frac{F^3}{\mu_T} \right)^{1/2} \frac{\partial F}
{\partial \phi} \frac{d\phi}{\phi}
\end{equation}
the grain velocity $V_{heap}(0)$ at the free surface
\begin{equation}\label{vheap}
\frac{V_{heap}(0)}{\sqrt{gD}}=\frac{( {\rm sin}(\theta)-\mu  {\rm cos}(\theta))^{1/2}}
{\left(1-\frac{\mu_N}{\mu_T}(\tan(\theta)-\mu)\right)^{3/2}} 
\int_{\phi_m}^{\phi_M} \left(\frac{F}{\mu_T} \right)^{1/2} \frac{\partial F}
{\partial \phi} \frac{d\phi}{\phi}
\end{equation}
and the relative velocity profile
\begin{equation}\label{vprof}
\frac{V_{heap}(z)}{V_{heap}(0)}=\frac{\int_{\phi_{heap}(z)}^{\phi_M}\left
(\frac{F}{\mu_T} \right)^{1/2} \frac{\partial F}{\partial \phi} \frac{d\phi}
{\phi}}{\int_{\phi_m}^{\phi_M} \left(\frac{F}{\mu_T} \right)^{1/2}
\frac{\partial F}{\partial \phi} \frac{d\phi}{\phi}}
\end{equation}
Since the free-surface velocity and the flux are expected to have finite
values for $ \theta_{min} < \theta <\theta_{max} $, the two functions
$ F(\phi)$ and $\mu_T(\phi)$ must be such as to guarantee the convergence
of the above integrals. In this case $V_{heap}(0)$ and $Q_{heap}$ are function
of $\theta$ with numerical prefactors depending on one's peculiar choice for
$F$ and $\mu_T$. In what follows we adopt the simple expressions
\begin{equation} 
F=F_0 {\rm Log}\left(\frac{\phi_M-\phi_m}{\phi_M-\phi} \right) \,\,
{\rm and} \,\, \mu_T=\mu_{T0} \left( \frac{\phi_M-\phi_m}{\phi_M-\phi} \right)^2.
\label{eq:FmuT}
\end{equation}
The same expression for $F$ was already proposed by Savage \cite{savage82,savage98} and leads to an
exponential-like volume fraction profile:
\begin{equation}
\phi_{heap}(z,\theta)= \frac{\phi_M}{1+(\frac{\phi_M}{\phi_m}-1)e^{-z/L(\theta)}}
\label{plog}
\end{equation}
with
\begin{equation}
L(\theta)= \frac{F_0D}{\phi_M(1-\frac{\mu_N}{\mu_T}(\tan(\theta)-\mu))}
\label{ltetha}
\end{equation}
$L(\theta)$ represents the typical thickness of the layer flowing down the heap, and it increases from 
$L(\theta_{min})=\frac{F_0D}{\phi_M}$ to infinity when $\theta=
\theta_{max}$. The relative velocity profile is also exponential-like for 
$\frac{z}{L(\theta)}   \stackrel{>}{\sim} 2$ (see figure \ref{fig:fig2}) but displays a Bagnold-like region of inverse 
concavity for $ \frac{z}{L(\theta)}  \stackrel{<}{\sim} 0.2$ (see figure \ref{fig:fig3}).
In fact, the numerical solution can be fitted by the analytical 
expression
\begin{equation}\label{vsimpl}
1 - \frac{V_{heap}(z)}{V_{heap}(0)}= \left ( \frac{\phi_{heap}(z)-\phi_m}
{\phi_M-\phi_m} \right)^\frac{3}{2}=
 \left ( \frac{1-e^{-z/L(\theta)}}{1+ (\frac{\phi_M}{\phi_m}-1)  e^{-z/L(\theta)}} \right)^\frac{3}{2}.
\end{equation}
With $\phi_m = 0.5$ and $\phi_M = 0.65$ the total flux flowing down the heap is 
$$
\frac{Q_{heap}}{D\sqrt{gD}}=1.4 
\frac{F_{0}^{5/2}}{\mu_{T0}^{1/2}}
 \frac{( {\rm sin}(\theta)-\mu  {\rm cos}(\theta))^{1/2}}
{\left(1-\frac{\mu_N}{\mu_T}(\tan(\theta)-\mu)\right)^{5/2}}.
$$
The dependence on $\theta$ of $L$ and $Q_{heap}$ are represented in Fig. \ref{fig:fig4}, with $\mu =0.36$ and $\mu_N/\mu_T = 4.7$.   

\subsection{Chute on rough plates}

The flows over inclined rough plates are more difficult to handle because the constitutive
equations (\ref{tzz}) and (\ref{txz}) hold only in the bulk of the dense 
granular 
medium and are likely to be modified close to the rough plate. 
Since the role of the plate rugosity is difficult to assess quantitatively,
we discard the description of the "basal layer" close to the plate \cite{ancey,louge} and 
assume a slip velocity $V_s$ at some distance $\delta$ above the rough plate.
Then we apply (\ref{tzz})
and (\ref{txz}) to a layer of thickness $h$, so that the free surface is located
at a distance $h+\delta$ above the rough incline. In the slab of thickness $h$,
the solid fraction 
increases from $ \phi_m$ at the free surface to the value $ \phi_{heap}(h)$
at a distance $\delta$ from the rough plate where the velocity is $V_s$. 
The total flux through the core region is
now given by:
{\setlength\arraycolsep{2pt}
\begin{eqnarray}
\frac{Q_{plate}}{D\sqrt{gD}}&=&\frac{F(\phi_{heap}(h))}{1 - \frac{\mu_N}{\mu_T}
(\tan(\theta)-\mu)} \frac{V_s}{\sqrt{gD}} +
\nonumber\\
& & {}+\frac{( {\rm sin}(\theta)-\mu  {\rm cos}(\theta))^{1/2}}
{\left(1-\frac{\mu_N}{\mu_T}(\tan(\theta)-\mu)\right)^{5/2}}
\int_{\phi_m}^{\phi_{heap}(h)} \left(\frac{F^3}{\mu_T} \right)^{1/2} \frac{\partial F}
{\partial \phi} \frac{d\phi}{\phi}{}.
\label{qplate}
\end{eqnarray}}
A rough plate is likely to slow down the core region more efficiently than a
heap would do and we expect $ V_s \le V_{heap}(h)$. As a consequence, 
$Q_{plate}(h,\theta)$ as given in (\ref{qplate}) is not expected to exceed 
$Q_{heap}(\theta)$ given in (\ref{qheap}). When forcing a flux $Q_{plate}$ to
flow down a rough plane inclined at angle $\theta$, two different situations
are encountered: when $ Q_{plate}$ is larger than $ Q_{heap}(\theta)$, 
the granular
medium will rearrange so as to flow down over a heap of angle $\theta+\alpha$
with $ Q_{plate}=Q_{heap}(\theta+\alpha)$. This gives a possible
explanation for the ``immature sliding flows'' that were observed in some 
experiments \cite{anceve,savage82}. Due to the very large increase of $Q_{heap}$ with $\theta$
(see figure (\ref{fig:fig4})) and because the experimental flux is limited
to some maximal value, immature sliding flows were observed for small angles
close to $\theta_{min}$ only.
Conversely, when $Q_{plate}$ is smaller than $Q_{heap}(\theta)$ the whole 
layer of thickness $h$ is in motion
with a velocity everywhere larger than $V_s$.
Moreover, when $h/L(\theta) \stackrel{<}{\sim}  0.2$, the Bagnold-like velocity 
profile (which could hardly be observed in heap flows, see fig. (\ref{fig:fig3}))
is now invading the whole core region. In fact, when expressions (\ref{eq:FmuT}) are taken for granted and $h/L(\theta) \stackrel{<}{\sim} 0.2$, the total flux (\ref{qplate})
has the special form
\begin{equation}\label{qsimp}
\frac{Q_{plate}}{D\sqrt{gD}}=\frac{\phi_m V_s h}{D \sqrt{gD}}+
\frac25 \phi_m^{3/2} \left(\frac{( {\rm sin}(\theta)-\mu  {\rm cos}(\theta))}
{\mu_{T0}}\right)^{1/2}  \left(\frac{h}{D} \right)^{5/2}.
\end{equation}
When the role of the velocity slip can be neglected, the second
contribution gives the $h^{5/2}$ scaling which seems to be corroborated by 
experiments \cite{silbert}, as well as numerical simulations \cite{pouli99}.

\section{Confined shear flow }\label{sex:con}

In the two-dimensional shear flows we will consider the pressure load exerted on the boundaries of the granular medium is supposed to be applied along direction $z$, which is thus the direction of main compression. Because gravity plays a minor role concerning the compressive forces, the constitutive relation for $\tau_{zz}$ is simply (compare with (\ref{tzz}))
\begin{equation}
\tau_{zz}  =   \rho D^2 \mu_N(\phi) \left(\frac{dV}{dz} \right)^2  +\rho g D F(\phi),
\label{tzzbis}
\end{equation}
\noindent
whatever the angle $\theta$ between the $z$ axis and gravity. The flow is along the $x$ axis and the constitutive relation for the shear stress $\tau_{xz}$ is still given by (\ref{txz}), without any change as compared to the free-surface case.
 
\subsection{Plane shear flow}
\begin{figure}
\centerline {  
\includegraphics[width=10cm]{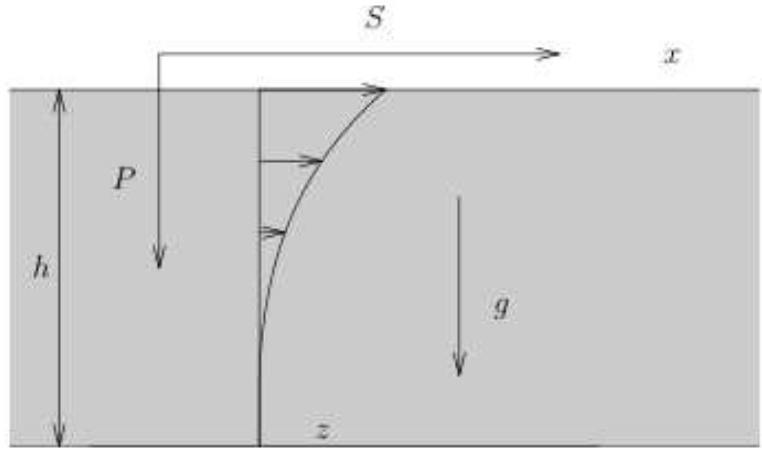}}
\caption{
Planar confined shear flow of an 
infinite horizontal granular layer bounded by two plates separated by a fixed 
distance $h$. The pressure load $P$ and
the gravity are both oriented along the direction $z$, the shear stress $S$ and the flow are along direction $x$).
 } 
\label{fig:fig5}
\end{figure}

As a first type of confined shear flow, we consider the planar shear of an 
infinite horizontal granular layer bounded by two plates separated by a fixed 
distance $h$. The pressure load and
the gravity are both oriented along the direction $z$ and the flow is along 
direction $x$ (see figure \ref{fig:fig5}). The equations of motion result in a constant 
shear stress $S$ and a variable normal stress:
$$ \tau_{xz}=S \;\; {\rm and} \;\; \tau_{zz}(z)=P(0)+\rho g \int_0^z \phi(\xi)
d\xi $$
where $P(0)$ is the pressure load exerted on the upper plate $z=0$ ($z=h$ stands for
the lower plate). We will distinguish the situation without and with gravity,
the first case corresponding to numerical simulations and the second one to experiments.  

\subsubsection{Without gravity}

In this case the normal stress is also a constant $P$ all over the granular 
layer and the constitutive equations (\ref{tzzbis}) and (\ref{txz}) give:
\begin{eqnarray}
\rho D^2 \mu_T(\phi) \left( \frac{\partial V}{\partial z} \right)^2=S-\mu(\phi)P \nonumber \\
\rho D^2 \mu_N(\phi) \left( \frac{\partial V}{\partial z} \right)^2=P-\rho g D F(\phi)
\end{eqnarray}
Depending on the sign of $S -\mu(\phi)P$, we will have a static or a moving 
medium. In the static case the pressure load is noted $P_0$ and the shear is 
such that $S \le \mu(\phi_0)P_0$ where $\phi_0$ is the constant compaction of
the medium related to the pressure load through $ P_0=\rho g D F(\phi_0)$. 
In the dynamic case, the compaction is still $\phi_0$ (mass conservation) 
while the shear $S$ is now larger than $\mu(\phi_0) P_0$. The
velocity gradient is constant:
$$ \rho D^2 \left( \frac{\partial V}{\partial z} \right)^2 =
\frac{S-\mu(\phi_0)P_0}{\mu_T(\phi_0)+\mu(\phi_0) \mu_N(\phi_0)} $$
Due to dilatancy effects the pressure load exerted on the plates is 
necessarily larger than in the static case, following:
$$ P(S)=P_0+\frac{S-\mu(\phi_0)P_0}{\mu(\phi_0)+\mu_T(\phi_0)/\mu_N(\phi_0)}.$$
As a consequence the effective friction coefficient is a function $\phi_0$ and $P$:
$$
\frac{S}{P}= \mu(\phi_0) + \frac{\mu_T(\phi_0)}{\mu_N(\phi_0)} (1 - \frac{\rho g D}{P}F(\phi_0)).
$$ 
\subsubsection{With gravity}

In this case the normal stress increases in the downward direction so that the
constitutive equation (\ref{txz}) results in 

$$ \rho D^2 \mu_T(\phi) \left( \frac{\partial V}{\partial z} \right)^2=S-\mu(\phi)P(0)-\mu(\phi) \rho g \int_0^z \phi(\xi)d\xi $$
It is then clear that the gravity-induced extra compaction possibly induces 
shear localization because the right-hand side can have a different sign
in different parts of the flow. To 
simplify this issue we will now introduce the same assumptions we have 
previously used in the free-surface shear flows, namely that $\mu$ and
$\mu_T/\mu_N$ do not depend on $\phi$ while $F(\phi)$ and $\mu_T(\phi)$
are given by (\ref{eq:FmuT}). We will first describe the static case before
considering grain motions. Because the compaction on the upper plate is
nesessarily different in the static and the dynamic cases, we define 
$P_0(0)$ as the pressure load exerted on the upper plate when the granular 
medium is motionless and $ \phi_0(z)$ as the static compaction profile. 
As long as $ S \le \mu P_0(0)$, the granular slab is motionless, the 
compaction $\phi_0(0)$ at the upper plate satisfies 
$P_0(0)=\rho g D F(\phi_0(0))$ and the compaction profile is:
$$ \phi_0(z)= \frac{\phi_M}{1+\left(\frac{\phi_M}{\phi_0(0)}-1 \right)e^{-z/L_0}} \;\; {\rm with} \;\; L_0=\frac{F_0}{\phi_M}D $$
When the granular medium is flowing, the compaction profile $ \phi(z) $ 
displays larger gradients and becomes
$$ \phi(z)= \frac{\phi_M}{1+\left(\frac{\phi_M}{\phi(0)}-1 \right)e^{-z/L}} \;\; {\rm with} \;\; L=\frac{L_0}{1+\mu \frac{\mu_N}{\mu_T}} $$
where $ \phi(0)$ is the new compaction at the upper plate. Since mass 
conservation requires
$$ \int_0^h [\phi(z)-\phi_0(z)]dz=0, $$
it is clear that $ L< L_0$ results in $ \phi(0) < \phi_0(0)$ and $ \phi(h) 
> \phi_0(h) $. The compaction of the moving medium is thus reduced at the
upper plate as compared to its static value while it is enhanced at the lower
plate.

\begin{figure}
\centerline {  
\includegraphics[width=15cm]{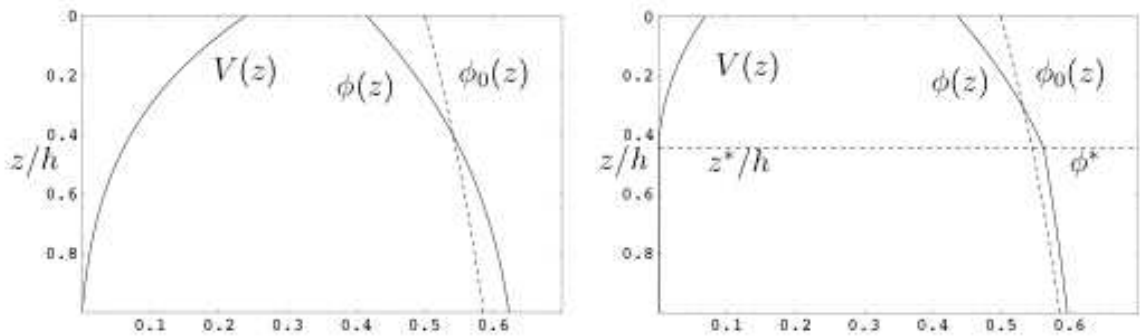}}
\caption{Velocity profile $V(z)$ and compaction profile $\phi(z)$, left large $S$, right small $S$. 
} 
\label{fig:fig6}
\end{figure}

The velocity profile is then deduced from the compaction profile
$$ \left(1+\mu \frac{\mu_N}{\mu_T} \right) \frac{D}{g} \left( 
\frac{\partial V}{\partial z} \right)^2= \frac{S^*-\mu F(\phi)}{\mu_T(\phi)} $$
where $S^*$ is the dimensionless shear $ \frac{S}{\rho g D}$. Let us define
the volume fraction $ \phi^*$ such that $ S^*= \mu F(\phi^*) $. It is clear
that $ \phi^* > \phi_0(0)$ because $S> \mu P_0(0)$. However the above equation
implies  that motion exists for compactions less than $\phi^*$ only. This 
condition leads to check the self-consistency relation $ \phi(z) < \phi^*$ 
for $0 < z < h$. This condition is automatically satisfied in the upper 
part of the flow since $\phi(0)<\phi_0(0) < \phi^*$. But it may be not in the 
lower part, thus leading to a shear localization. This localization
phenomenon is here depending on $ S^*$ and $h/L$. Fig (\ref{fig:fig6}) shows  the compaction and velocity profiles for two different values of $\phi^*(S^*)$ 
investigating the two different situations depending whether localization
occurs or not.

\subsection{Vertical chute flows}

\begin{figure}
\centerline {  
\includegraphics[width=10cm]{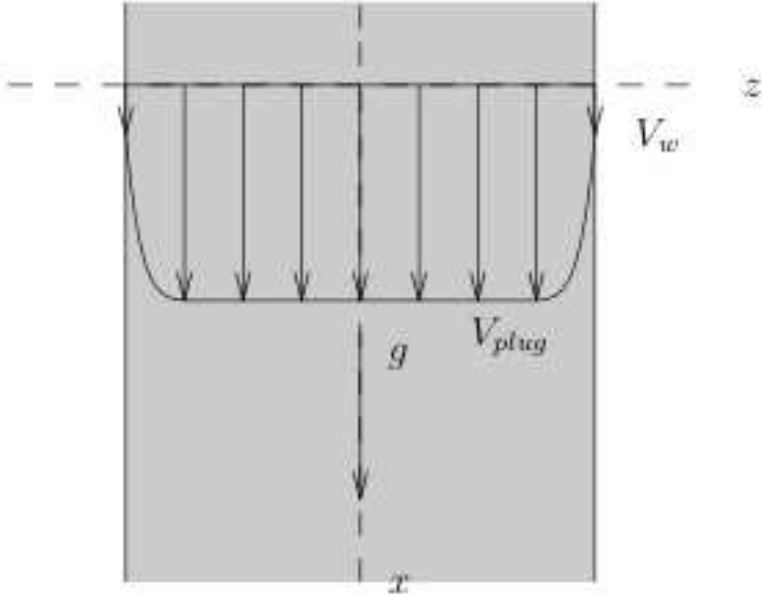}}
\caption{ 
Vertical chute  flow between two  rough plates. } 
\label{fig:fig7}
\end{figure}

\begin{figure}
\centerline { 
\ \includegraphics[width=10cm]{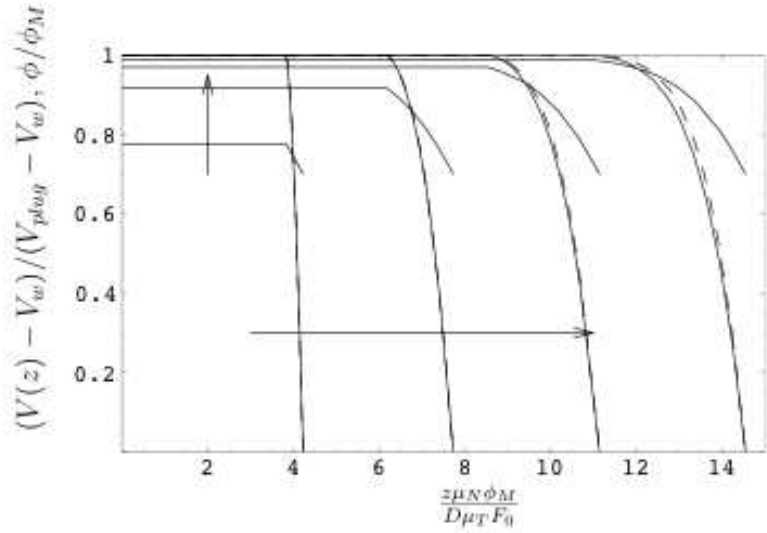}}
\caption{ Reduced velocity profiles $ (V(z) - V_w)/(V_{plug}-V_w) $ versus the adimensional distance
 $\frac {z \mu_N\phi_M}{D \mu_T F_0}$ to the center. 
The arrow is in the increasing values of $\frac{P}{\rho g D F_0}$.
The dashed curve represents the approximate expression (\ref{eq:fit}). 
The reduced compaction profiles $ \phi/\phi_M$ with $ \phi_w/\phi_M=0.7$ are plotted as well.
} 
\label{fig:fig8}
\end{figure}

A second type of confined shear flow is the chute between two vertical plates
(see figure \ref{fig:fig7}). The compaction is due to a pressure load $P$ exerted on the two plates along direction $z$. The flow and the gravity are oriented along 
direction $x$. The equations of motion result in a constant normal stress and
a variable shear stress, by contrast to the preceding case:
$$ \tau_{zz}=P \;\; {\rm and} \;\; \tau_{xz}=\rho g \int_0^z \phi(\xi)d\xi, $$
where $z=0$ corresponds to the symmetry plane located between the two plates at
which the shear stress vanishes. The constitutive relation (\ref{txz}) implies

$$ \rho D^2 \mu_T(\phi) \left(\frac{\partial V}{\partial z} \right)^2=
\rho g \int_0^z \phi(\xi)d\xi- \mu(\phi)P.$$
Either the right-hand side is everywhere negative (due to a very high pressure load) and
the medium is motionless or there is a central region of the flow in which the shear stress does not
exceed $\mu P$ and consequently where the strain rate vanishes. In this plug
flow regime the solid fraction is a constant $ \phi^*$ related to the pressure
load as $P= \rho g D F(\phi^*)$. The thickness $z^*$ of the plug flow depends
on $ \phi^*$ (hence on the pressure load)
$$ \frac{z^*}{D}= \frac{\mu(\phi^*)}{\phi^*} F(\phi^*). $$
Close to the vertical plates, there is a shear layer where the velocity  decrease to $V_w$ dependent on the plate roughness. In this parietal shear layer, 
the constitutive equations (\ref{tzzbis}) and (\ref{txz}) imply:

\begin{equation}
\left( \frac{D}{g} \right)^{1/2} \frac{\partial V}{\partial z}= -\left(
\frac{F(\phi^*)-F(\phi)}{\mu_N(\phi)} \right)^{1/2}
\label{velochute}
\end{equation}
and
\begin{equation}
D \frac{\partial \phi}{\partial z}=\frac{\phi}{\frac{\partial}{\partial \phi}
\left[ (\mu+ \frac{\mu_T}{\mu_N})F(\phi^*)-\frac{\mu_T}{\mu_N} F(\phi) \right]}.
\label{denschute}
\end{equation}
To obtain more definite results we again consider the assumptions 
 already made  for gravity-driven and plane shear flows, namely that $\mu$ and
$\mu_N/\mu_T$ are independant of the solid fraction while $F(\phi)$ and 
$\mu_T(\phi)$ are given by (\ref{eq:FmuT}). Then, the compaction profile in the 
shear layer $z^* < z < z_w$ is:
\begin{equation}
 \phi(z)= \frac{\phi_M}{1+\left(\frac{\phi_M}{\phi^*}-1 \right) e^{\frac{z-z^*}{L^*}}}
\label{eq:phic}
\end{equation}
where $L^*$ is the typical shear layer thickness:
$$ \frac{L^*}{D}=\frac{\mu_T F_0}{\mu_N \phi_M}. $$
For the flow to be  dense  up to the vertical plates, the wall compaction
$\phi_w$ must be larger than $\phi_m$ and the shear layer thickness is
$$ \frac{z_w-z^*}{L^*}=Log\left(\frac{\frac{\phi_M}{\phi_w}-1}{\frac{\phi_M}
{\phi^*}-1} \right).$$
As a consequence, the distance $ 2 z_w$ between the two plates is a function of
$\phi^*$ (hence of $P$) and of $ \phi_w$ (hence of the plate roughness).
Concerning the velocity, it increases from a value $V_w$ at the wall to a 
value $V_{plug}$ in the central part. The computed relative velocity field
is represented in Fig (\ref{fig:fig8}) together with the fit
\begin{equation}
 \frac{V(z)-V_w}{V_{plug}-V_w}=1- \left(\frac{\phi^*-\phi(z)}{\phi^*-\phi_w}
\right)^{3/2}.
\label{eq:fit}
\end{equation}

\section{ Comparison with experimental and (or) numerical data}\label{sex:exp}
\subsection{ Plane shear flow}
Neglecting the influence of gravity (as was done in most numerical simulations) our model leads
to a uniform solid fraction and to a uniform velocity gradient, in conformity with results presented in figures 5b and
5c of \cite{midi} for the dense flow regime. When gravity is taken into account, a shear localization is possible,
depending on the magnitude of the pressure load as well as on the thickness of the granular layer. Unfortunately, we
are unaware of experimental or numerical data with which the predictions of Fig. (\ref{fig:fig6}) could be tested.

\subsection{Vertical chute flow }
The uniform solid fraction and the uniform velocity in the core region are correctly
reproduced by the model. Concerning the sheared regions closed to the vertical boundaries, the relative velocity
profile (\ref{eq:fit}) and the compaction profile (\ref{eq:phic}) are quite similar to results presented in fig. 7b and 7c of \cite{midi}.

\subsection{Heap flow}
The solid fraction profile (\ref{plog}) and the velocity profile (\ref{vsimpl}) are quite close to those represented in
fig.9b and 9c of \cite{midi} and in fig.9a and 9b of \cite{rach}. In particular, the velocity profile displays a
Bagnold-like profile very close to the free-surface ($z < 0.2 L(\theta)$), a quasi-linear profile in the central part of
 the flow ($0.2 < z/L(\theta) < 2$) and finally an exponential tail for the deepest parts of the flow, observed in \cite{koma}.
At variance with confined flows for which the shear was localized in boundary layers with thickness of the order of a
few grain diameters, heap flows are characterized by a thickness $L(\theta)$ of a few grain diameters when $\theta$ is
slightly larger than $\theta_{min}$ but which increases to quite large values when $\theta$ is close to $\theta_{max}$. A similar unlimited increase of the grain flux $Q_{heap}$ is observed for $\theta$ close to $\theta_{max}$, as seen in fig. \ref{fig:fig4}. Such a behaviour is difficult to observe experimentally due to the limited values of $Q_{heap}$ that can be achieved in usual laboratory devices. According to (\ref{qheap}) and (\ref{ltetha}), our model predicts $L \sim (Q_{heap})^{2/5}$ when $\theta$ is not close to $\theta_{min}$,
a result slightly different from the scaling $L \sim (Q_{heap})^{1/2}$ suggested by fig.9j of  \cite{midi}.

\subsection{Rough inclined planes}
We explained the appearance of the so-called "immature sliding flows": they develop when the imposed flux $Q_{plate}$ over a plate with inclination $\theta$ is larger than the flux $Q_{heap}(\theta)$ which would fall down a heap
 with the same slope. Since $Q_{plate}$ is experimentally limited to some maximum value, immature flows are observed for $\theta$ close to $\theta_{min}$ only. When $\theta$ comes close to $\theta_{max}$, the thickness $h$ of the granular layer flowing over the rough incline is smaller than the thickness of the grain layer which would flow down a heap with similar slope. And when $h$ is less than $0.2 L(\theta)$, the Bagnold-like velocity profile is invading the whole flowing layer, with the
$h^{5/2}$ scaling for the flux $Q_{plate}$ as a direct consequence (provided the first contribution to 
(\ref{qsimp}) is negligible). 
 The main drawback of our model is its inability to explain the quantity $h_{stop}(\theta)$ introduced by Pouliquen \cite{pouli99} and which was confirmed in numerical simulations 
\cite{silbert}. The first reason is that we assumed the friction coefficient $\mu$ to be independent of the solid fraction. As a consequence $\theta_{min}$ is a constant and $h_{stop}$ vanishes as soon as $\theta > \theta_{min}$. A second reason is the possible inadequacy of our model close to the rough incline. In this basal or frictional layer \cite{ancey,louge}, the particle rotation plays an important role, the grain stress tensor is possibly non-symmetric and the solid fraction has a perturbed profile. All these phenomena would require a specific modelling. In fact the explanation of $h_{stop}(\theta)$ proposed by Mills et al. \cite{ mills99} involves constitutive relations which are different close to the boundaries from those holding in the bulk.

\subsection{Annular shear}
This special kind of shear flow was not considered here because to describe it, we would need to give a constitutive equation for the $\tau_{xx}$ component of the granular stress, besides those for $\tau_{xz}$ and $\tau_{zz}$. This will be done in future work.

\section{Conclusion}

We proposed a model for dense granular flows which considers the solid fraction as the main microstructural parameter. The granular stress is partitioned in a way similar to that proposed by Savage \cite{savage98,savage82}. One of the distinctive features is a completely explicit expression for the contact stress which involves a function $F(\phi)$ of the solid fraction. The solid fraction profile mainly depends on the compressibility $dF/d\phi$ while the velocity profile is bound to $F(\phi)/\mu_T(\phi)$ where $\mu_T(\phi)$ is somehow analoguous to the effective viscosity used in the kinetic theory approach of dilute granular flows \cite{bocquet,louge}. In principle the complete model contains two more functions of the solid fraction ($\mu(\phi)$ and $\mu_N(\phi)$) but we strived to show that not so bad predictions could be obtained after assuming the friction coefficient $\mu$ and the dilatancy ratio $\mu_N/\mu_T$ to be independent of the solid fraction. Obviously, these are simplifying assumptions which can be released and improved. We also checked that the tentative (and simple) expressions (\ref{eq:FmuT}) for $F(\phi)$ and $\mu_T(\phi)$ led to sound predictions. Needless to say that these expressions also can be improved.
   The main drawback of constitutive equations (\ref{tzz}) and (\ref{txz}) is their possible failure in a thin layer close to rough boundaries. Their main advantage is to contain all the ingredients necessary to interpret the shear-localization phenomenon, and to be able to explain the quite different velocity profiles appearing in the stationary shear flows of dense granular materials.

\clearpage

\end{document}